\documentclass[prl,twocolumn,superscriptaddress,showpacs]{revtex4-1}
\usepackage{color}
\usepackage{amsmath}
\usepackage{graphicx}
\usepackage{amssymb}
\usepackage{amsthm, amscd}
\usepackage{amsfonts}
\usepackage{times}
\usepackage{wasysym}

\begin{document}

\title{Telegraph noise and the  Fabry-Perot quantum Hall interferometer}

\author{ B. Rosenow}\affiliation{Institut f\"ur Theoretische Physik, Universit\"at Leipzig, D-04103, Leipzig, Germany}
\author{Steven H. Simon}
\affiliation{Rudolf Peierls Centre for Theoretical Physics, University of Oxford,OX1 3NP, United Kingdom}

\date{\today}

\begin{abstract}
We consider signatures of abelian and nonabelian quasiparticle statistics in quantum Hall Fabry-Perot interferometers.
When quasiparticles enter and exit the interference cell, for instance  due to glassy motion in the dopant layer,
the anyonic phase can be observed in phase jumps.  In the case of the nonabelian $\nu=5/2$ state, if the interferometer is small, we argue that  free Majoranas in the interference cell  are either strongly coupled to one another  or are strongly coupled to the edge.  We analyze the expected phase jumps and in particular suggest that changes in the fermionic parity of the ground state should gives rise to characteristic jumps of $\pi$ in the interference phase.
\end{abstract}


\maketitle

The search for low-energy Majorana fermions  has become a focus of a both condensed matter physics and quantum information \cite{PhysicsToday,review}.   The experimental system that appears closest to finding evidence for the existence of Majoranas is the $\nu=5/2$ quantum Hall state,  where recent experiments
have found evidence  for   excitations with one quarter of the electron charge \cite{Radu+08,Dolev+09,Vivek+11}, which are expected if
the $\nu = 5/2$ state is due to pairing of electrons. From a theoretical point of view,  the most likely
description is the Moore-Read Pfaffian \cite{MR}  state (or its particle-hole conjugate \cite{apf}, which for the purpose of this paper is identical)
with charge $e/4$ excitations that host Majorana fermions and have non-abelian statistics in the Ising universality class\cite{commentonising,Parsainsistsonthisreference,Imaddingthisreftomakeapoint,Baraban,Storni},  and thus have the potential for topological quantum computation \cite{review}.

The nonabelian statistics of $\nu=5/2$  quasi-particles (QPs) was theoretically predicted to be observable in
Fabry-Perot interference experiments \cite{fradkin98,dassarma05,stern06,bonderson06}. Under ideal conditions where all localized QPs are far apart from each other and far away from the edge of the interferometer, the non-abelian statistics manifests itself in the even-odd effect \cite{stern06,bonderson06}: the fundamental  harmonic of the interference signal disappears when an odd number of QPs is inside the interferometer cell. For an even number of QPs in the interior of the interferometer, there are degenerate
states due to the Majorana zero modes in the interference cell, and the interference phase may switch by $\pi$ depending on the parity of the fermion number within the cell. For experimental
realizations of Fabry-Perot interferometers in micron scale structures, such as the one  by Willett et al.~\cite{Willett}
and by Kang et al.~\cite{Kang},
the idealized picture  as described above may not apply. The goal of this paper is to consider a possible set of experimental conditions which could potentially be in agreement with the behavior of the experiments. One of the key features we will focus on is the effect of telegraph noise\cite{Grosfeld}  --- to what extent it can influence the measurement, and how it can be used as a tool for extracting physical information from the measurements.

At integer filling fraction and in the absence of interaction effects,  the addition of a single quasiparticle (an electron or hole) to the cavity does not change the interference phase (since an electron on the edge encircling an electron or hole in the interior of the cavity accumulates a total phase of $\pm 2 \pi$).    However, at fractional filling $\nu$ where quasiparticles have anyonic statistics (let us say, abelian statistics for now) a quasiparticle on the edge encircling a quasiparticle or quasihole in the interior accumulates a phase which is a fractional multiple of $2 \pi$.   Thus, if a quasiparticle or quasihole enters the cavity, it changes the phase of the interference.   Typically, the conductance  will be of the form (again assuming abelian statistics)
\begin{equation}
\label{eq:telonethird}
G = G_0 + G_1 \cos(\theta)
\end{equation}
where
%
\begin{equation}
   \theta   =  2 \pi e^*(\phi + \beta V_G) + N_L \theta_a  + \theta_C (\phi, V_G, N_L)
   \label{phase.eq.1}
\end{equation}
Here, $\phi$ is the (dimensionless) flux through a reference area $A_0$ for the interferometer,  $e^*$ is the charge of quasiparticles in units of the electron charge, $N_L$ is the number of quasiparticles inside the interference loop, which may change abruptly, and $\theta_a$ is the anyonic phase which is $2 \pi/3$ for $\nu=1/3$ (or $4/3$ or $7/3$ etc.). $V_G$ denotes a change in side-gate voltage measured relative to a reference value, and  the coupling of voltage to the interferometer area is described by the parameter $\beta = {B \over \phi_0} {\partial A_0 \over \partial V_G}$.
Here $\theta_C(\phi, V_G, N_L)$ is the Coulomb correction to the interference phase.      Ideally, one would like to  observe the anyonic phase $\theta_a$ directly.

The early theoretical discussions of the quantum Hall Fabry-Perot interferometer \cite{dassarma05,stern06,bonderson06,chamon97,fradkin98}  neglected the strong Coulomb interaction
(and hence the correction $\theta_C$)
that can occur in a pinched-off Fabry-Perot cavity, and focused on the physics deep in the so-called Aharonov-Bohm (AB) regime where $\theta_C$ is small.
 However, more recent theoretical work \cite{RosenowHalperin,Nederetal} supported by several experiments\cite{Godfrey,YimingZhang,Ofek} showed that a different regime where the strong Coulomb interaction dominates the physics (the so-called ``Coulomb Dominated (CD) Regime") more typically occurs.
Thus, it may be necessary  to disentangle the anyonic phase $\theta_a$ from the Coulomb correction.

Fluctuations in the number of localized quasiparticles, $N_L$, can have different origins. One possible
source of fluctuations in $N_L$ are voltage fluctuations caused by the glassy dynamics of charges in the
donor layer,  which can be very slow. We will focus on this effect, and we will consider two models of how
this may result in noise measured in the conductance.

In our first model, we consider the noise to be equivalent to a random change in the gate voltage.   This gate voltage may attract discrete quasiparticles into the interferometer.   Far in the AB regime, when Coulomb effects are weak, this should result in phase slips in the interference pattern given by the anyonic phase $\theta_a$ whenever a quasiparticle is added to the interferometer.   However, if Coulomb effects are stronger, there can be a deviation $\delta \theta_C$ from this ideal value.   To remain in the AB regime, the Coulomb effects must not be too large, and one can bound the magnitude of $\delta \theta_C$ within the AB regime.  The derivation is straightforward and is given in Supplemental Material A, with the result that (at $\nu=1/3, 4/3, 7/3$) the Coulomb contribution $\delta \theta_C$ to the phase slip is always negative, with its magnitude at most half as big as the anyonic contribution $\theta_a$.    In the CD regime, however, the Coulomb correction can be as big as $\theta_a$ in magnitude.

We now consider a second model where the fluctuations in the donor layer are strongly coupled to the Coulomb charge of the interferometer.  For simplicity we consider two different configuration of donor impurities -- and correspondingly two possible values of $N_L$.    In this model fluctuations between the two states occurs only when there is a near degeneracy of the energy of the two states.   If the fluctuations in the donor layer occur physically close to the 2DEG in the interferometer, we can have a situation where the Coulomb correction $\delta \theta_C$ to the ideal phase slip is very small.  A detailed calculation is provided in Supplemental Material B.   In essence the charging effect of the fluctuation of charge in the donor layer is roughly canceled by the addition of the quasiparticle charge  -- and this cancelation is enforced by the requirement that the two possible states of the system are energetically degenerate. Thus one may measure the ideal phase jump value even deep into the CD regime.

Experimentally, by examining the phase jumps that occur in the telegraph noise as the side gate voltage
is changed smoothly, one can attempt to measure the statistical parameter $\theta_a$.    Recent experiments at $\nu=7/3$ by Kang~\cite{Kang} have made precisely this type of measurement
and have observed a phase jump in agreement with the ideal anyonic phase  $\theta_a$.  This result
can be explained  if either the system is deep into the AB regime (which is unlikely for a small device) or the above described screening cancelation of our second model is being realized.

\begin{figure}[t]
\vspace*{-140pt}
\includegraphics[width=3.5in]{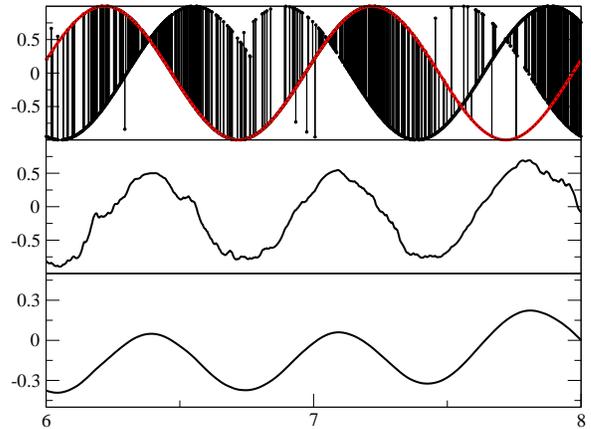}
\caption{A simulation of random telegraph noise during a scan of a $\nu=1/3$ (or 7/3) Fabry-Perot interferometer. Vertical axis is $\cos( 2 \pi t + 2 \pi N_L/3)$ where $t=  e^* \beta V_G$ is the horizontal axis which we can think of as time assume constant voltage sweep rate.  Here $N_L = \lfloor \eta t + \xi r \rfloor$ with $\lfloor x\rfloor$ denoting the  greatest integer smaller or equal to $x$, and $r$ being a gaussian random variable with unit variance.   The analytic results are then passed through a simulated lock-in amplifier with  an averaging time  $\tau$.  Here $\eta=1.25$ means roughly\cite{endnote1} 1.25 quasiparticles is added to the dot per cycle and $\xi=.3$ is the amplitude of the noise in units of quasiparticle number.  The correlation ``time" of the noise constant of the gaussian noise is $\tau_c=.001$  in the above units (Top) observed conductance with an averaging time shorter than the noise correlation time.  Red curve is a pure sine wave for reference (Middle) Observed conductance using an averaging time  $\tau = .0066$ larger than $\tau_c$.  Note that the detailed structure is hidden by the averaging.  (Bottom) Observed conductance using a very long averaging time   $\tau = .066$.   Note that the period of oscillation is distorted towards smaller values  as discussed in the text. }
\label{fig:simulatedthird}
\end{figure}

In Figure \ref{fig:simulatedthird} (top)  a simulated data set is displayed for $\nu=7/3$. The plotted conductance is given by Eq. \ref{eq:telonethird} where the variable $t=e^* \beta V_G$ is varied smoothly and $N_L$ is the integer part of a constant times $t$ plus a random component with a correlation ``time"
of $\tau_c$. This simulated data looks qualitatively similar to the experiments of Ref.~\cite{Kang}.
If one were to ``time average" over this telegraph noise, a very different signal could result as shown in Figure \ref{fig:simulatedthird} (middle and bottom).  Here,
the time averaging is achieved by Fourier transforming $t$ to $\omega$, multiplying the transform by
$e^{- \omega \tau}$ and inverting the transform. This procedure simulates a lock-in amplifier with time constant $\tau$.
In the limit where $\tau \gg \tau_c$ the observed $G$ in Eq. \ref{eq:telonethird} is replaced by its average over $N_L$ for each value of $V_G$.  In this case the observed period in $t$ will shift from $1$ to $1 / (1 + \frac{\theta_a}{2 \pi} \langle d\langle N_L \rangle/dt\rangle)$.   In Figure \ref{fig:simulatedthird} we show the drastic effect of a long measurement time constant if the telegraph noise is fast.  It should be understood that the model for random quasiparticle motion in the dot (explained in the caption) is quite crude.  Nonetheless it gives a feel for the physics.

We now turn to study the situation at $\nu=5/2$.  For a moment, let us assume all of the quasiparticles are stationary (no telegraph noise) and all of the quasiparticles in the cavity are far from each other and far from the edge.  We will also focus here on intereference of the nonabelian $e/4$ particles traveling around the cavity and ignore the abelian $e/2$ particles for the moment.   If there are an odd number of quasiparticles in the cavity, no interference should be observed\cite{stern06,bonderson06}.   If there is an even number of quasiparticles in the cavity, there are degenerate zero modes (or qubits) due to the nonabelian degrees of freedom associated with the quasiparticles.   Interference should be seen, but  the interference phase may be switched  by $\pi$ depending on the quantum number of the nonabelian degrees of freedom within the cavity (i.e., the setting of the nonabelian qubits in the cavity).    However, if the Majoranas in the bulk of the cavity are  coupled to the edge modes, then the qubits in the cavity can flip and, if the measurement time scale is sufficiently long, the two possible (opposite) phases of signals are both seen equally, resulting in cancelation of the interference signal\cite{Gangoffour}.   The rate at which the qubit is expected to flip is determined by the bulk-edge coupling (set by the distance from the bulk Majorana to the edge).  While this coupling is expected to decay exponentially with distance, given that the cavities in the experiments of Refs.~\onlinecite{Willett,Kang} are extremely small, an estimate of the decay length given by Ref.~\cite{Baraban} suggests that for a measurement time scale on the order of seconds,   the coupling of bulk to edge should always be sufficiently strong to destroy the interference signal\cite{Gangoffour}.   This raises the question of why interference should be seen at all.   There is, however, a plausible scenario which we will now discuss.

So far we have been assuming that the quasiparticles in the cavity are sufficiently far from each other that the nonabelian degrees of freedom are all zero energy.  However, again given that the cavity is small, the Majoranas couple to each other and the zero energy modes split.    If this splitting is larger than the temperature, then the qubits freeze into their lowest energy state and interference would again be seen for the case of an even number of quasiparticles in the cavity, but not for odd.  We thus need to consider the spectrum of the coupled Majoranas which depends on their detailed positions and therefore the detailed geometry of the dot.  We estimate\cite{SimonTalk} that there are roughly 20 quasiparticles in the dot and that they may be spaced on the order of 0.1 micron.   The spectrum should be  roughly given by energy levels equally spaced by
$\Delta E = t/N$ with $t$ the neighbor hopping possibly as large\cite{Baraban} as $200 \rm{mK}$ and $N$ is the number of quasiparticles.  This spacing of $\Delta E \approx 10 \rm{mK}$  is potentially large enough to allow interference to be seen at accessibly low temperatures\cite{endnote2}.  Although these numerical estimates are optimistic, they are not out of the question.

While this seems like a good explanation for why interference is seen at $\nu=5/2$, there is a complication again associated with the bulk-edge coupling.   The situation of having a bulk Majorana coupled to the edge has been discussed in detail in Refs.~\cite{BisharaNayak,Gangoffour,Wen} (See also \cite{Shtengel})  An important limit is when the Majorana is strongly coupled to the edge compared to the measurement voltage $e^*V$ (which experimentally is roughly on the scale of the temperature).  In this case the Majorana is absorbed into the edge --- and the situation becomes as if that particular Majorana were no longer in the cavity.     Considering again that the dot is very small and the bulk-edge coupling is likely to be substantial, this effect is one which we must address.

Unfortunately, determining the size of the bulk-edge coupling is even more uncertain, requiring detailed knowledge of the structure of the dot and the edge.   Attempts at electrostatic simulation\cite{SimonTalk} to determine positions of particles and edges suggests that it is not easily possible to have an excitation  gap higher than the temperature and yet always weak enough coupling to the edge that a lone Majorana will not be absorbed into the edge.    Instead, we assume the opposite inequality that the bulk edge coupling is larger (or on order of) $e^*V$. In this case, a lone Majorana is always absorbed into the edge and interference is observed when there are an odd number of quasiparticles in the dot as well as when there are an even number of quasiparticles in the dot (so long as the excitation gap is larger than or on order of $T$).   Note that if $e^* V$ is not much less than the bulk-edge coupling, then the Majorana is not very strongly coupled to the edge, and the amplitude of interference can be reduced and the phase slightly shifted \cite{Gangoffour,BisharaNayak}.

We next consider the phase of the interference pattern for both the $e/2$ or $e/4$ quasiparticles and how it may change if quasiparticles are hopping in and out of the dot --- analogous to what we considered for $\nu=1/3$ above.
For interference of $e/2$ particles traveling around the interferometer, the interference is given by Eq.~\ref{eq:telonethird} with $
N_L$ being the number of $e/4$ quasiparticles inside the interferometer (with an $e/2$ counting as two $e/4$'s) and $\theta_a = \pi/2$.  We will not study this case further since it is not very different from the above discussed $\nu=7/3$, and it is likely the tunneling of $e/2$ quasiparticles is less than that of of $e/4$ at any rate.

For interference of $e/4$ traveling around the interferometer, again assuming that any lone Majoranas are strongly coupled to the edge\cite{Gangoffour} and the remaining nonabelian modes are thermally frozen into a particular state, we find an interference pattern {\it also} of the form of Eq.~\ref{eq:telonethird} where $\theta_a=\pi/4$ (again with any $e/2$ quasiparticle in the interferometer counting as two $e/4$'s).   Thus deep in the AB regime we would expect to observe phase slips with an ideal value of $\pi/4$ due to quasiparticle addition.   Analogous to the case of $\nu=7/3$ above, we may consider two models of charge fluctuation from the donors.  For the first model, we can derive (see Supplemental Material A) that within the AB regime the maximum Coulomb correction to the ideal $\pi/4$ phase slip is negative like in the $\nu=7/3$ case, but with $|\delta \theta_c|=3 \pi/8$  quite large in magnitude compared to $\theta_a$.   In  the CD regime, the Coulomb correction can be up to $3 \pi \over 4$ in magnitude, three times larger than the statistical phase itself.  Within the second model (See Supplemental Material B) again we predict that the Coulomb correction can be quite small, even within the CD regime.

\begin{figure}[t]
\vspace*{-140pt}
\includegraphics[width=3.5in]{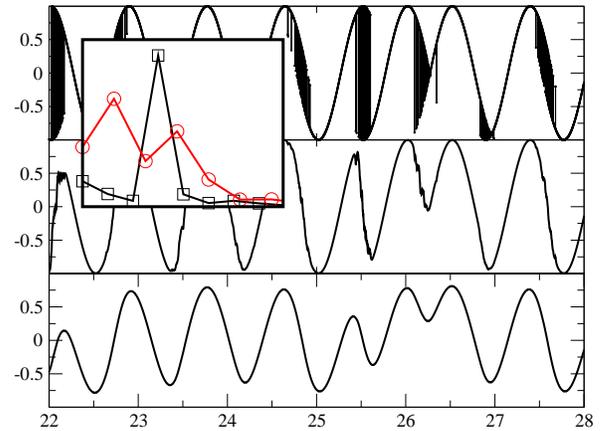}
\caption{Simulation of telegraph noise during a scan of a $\nu=5/2$ Fabry-Perot interferometer. Vertical axis is $\cos (2 \pi t + \pi f(N_L)/4)$ where $t= e^* \beta V_G$ is the horizontal axis and $N_L = \lfloor\eta t + \xi r \rfloor$ with $r$ again being a gaussian random variable of unit magnitude.   Here $f(n)$ is an integer valued function where $f(n+1) - f(n)$  takes the value $4$ with probability .3 (corresponding to a $\pi$ phase slip), and the value 1  with probability .5 (corresponding to $\pi/4$)  and and 5 with probability 0.2 ( corresponding to $5 \pi/4$ phase slips). For this plot the noise correlation time is $\tau_c=.001$, $\eta=1.45$ and $\gamma=.08$.  (Top) Results for an averaging time shorter than $\tau_c$.  The phase shifts of $\pi$, $\pi/4$ and $5 \pi/4$ are easily observed (Middle) Observed with an averaging time time $\tau=.0033$ these phase shifts are obscured.  (Bottom) Observed with an averaging time  $\tau=0.033$ the smooth behavior appears as if there are fast and slow periods.   (Inset) Windowed Fourier transforms of the bottom panel. (black squares) transform of region $t=22.5 - 25$ (red circles) transform of region between $t=25$ and $t=27$.
Qualitatively similar pictures are obtained with many different parameters used in the simulation.}
\label{fig:simulatedhalf}
\end{figure}

In addition to these phase slips due to quasiparticle addition, the interference pattern may be flipped (shifted by an additional phase of $\pi$) depending on the state of the frozen nonabelian degrees of freedom within the cavity. Indeed, each time the quasiparticles in the dot have their positions rearranged, this degree of freedom may be changed since the lower energy state of the qubit depends on the detailed configuration of quasiparticles\cite{Baraban}.    We should thus expect to see (ideally) phase slips of both $ \pi/4$ and $\pi$ where the slips of $\pi$ may occur concurrent with the slips of $\pi/4$ (resulting in $5 \pi/4$) or may occur separately\cite{endnote1}.   A rough simulation of this type of physics is shown in Figure \ref{fig:simulatedhalf} (top) and as above a low-pass filter (middle and bottom) is shown of the same data. We would like to mention that similarly to the case without bulk-edge and bulk-bulk coupling\cite{Stern+10}, the $331$-state can give rise to the same types of phase jumps.  Slips of $\pi$ are expected due to  spin flips in the bulk.  In the special case where only one of the two spin edge channels is interfering, then slips of $\pm \pi/4$ and $\pm 5 \pi/4$ are expected when the different spin quasiparticles enter or exit the dot.

{\em Relation to experiment:} the experimental results \cite{Kang} are in very good agreement with a model assuming strong coupling between both bulk Majorana degrees of freedom and strong bulk-edge coupling.
 The fact that the experimentally measured values of phase slips are described by the statistical phases without
 Coulomb correction can be explained by coupled interferometer/dopant model described in the supplemental material. With respect to the experiment of Ref.~\onlinecite{Willett}, it is possible that at least part of the non-sinusoidal behavior observed there  is a result of time averaging over phase jumps due to telegraph noise.   Unfortunately, it is extremely hard to estimate what the fundamental time scale of the telegraph noise should be (and it may differ from sample to sample) since it almost certainly is related to glassy behavior.  Furthermore, in Ref.~\onlinecite{Willett}, non-sinusoidal behavior is also observed at $\nu=2$ which, as mentioned above, does not have fractional phase shifts.   In this case, the non-sinusoidal behavior is most likely to be caused by Coulomb charging effects\cite{RosenowHalperin}.
 In the $\nu=5/2$ regime, it is possible that time averaging over $\pi$ phase slips  mimics a reduction of the
period of resistance oscillations, see Fig.~\ref{fig:simulatedhalf}.

{\em Acknowledgments:} We would like to thank W.~Kang for helpful discussions and for making his data available to us before publication, and acknowledge helpful discussions with  A.~Stern and B.I.~Halperin.  This work was supported in part by EPSRC grants EP/I031014/1 and EP/I032487/1, by the Aspen Center for Physics,  and by BMBF.

\widetext

\newpage

\subsection*{Supplemental Material A: Random Gate Voltage Model}
Following Refs. \onlinecite{RosenowHalperin,Nederetal}, we write the total phase in Eq.~\ref{eq:telonethird} as
\begin{equation}  \label{phase.eq.2}
 \theta   =  2 \pi e^*(\phi + \beta V_G) + N_L \theta_a
-   {K_{IL} \over K_I}    { 2 \pi e^* \over \Delta \nu}
\left( e^*  N_L  + \nu_{in} \phi - \gamma  V_G \right) \ \ ,
\end{equation}
where $\phi$ is the (dimensionless) flux through a reference area $A_0$, $e^*$ is the charge of quasiparticles in units of the electron charge, $N_L$ is the number of quasiparticles inside the interference loop, which may change abruptly, $\theta_a$ is the anyonic phase which is $2 \pi/3$ for $\nu=1/3$ or $7/3$, $V_G$ denotes a change in gate voltage measured relative to a reference value, and
$\Delta \nu = \nu_{\rm in} - \nu_{\rm out}$ is the difference in filling fraction between (fractional) quantum Hall states in the interior and exterior of the interferometer.   Here $K_{IL}$ parameterizes  the strength of the Coulomb coupling between localized charges in the interior of the interferometer and charges on the edge of the interferometer. $K_I$ is the charging energy for adding one electron to the edge surrounding the interference cell.  The change in the interferometer area due to the change in gate voltage is described by the parameter   $\beta = {B \over \phi_0} {\partial A_0 \over \partial V_G}$,
where we assume that the reference area changes smoothly as a function of gate voltage. The  coupling of the gate voltage to the localized charge in the interior of the interferometer is described by $\gamma = {\partial \overline{N} \over \partial V_G}$. Here, $\overline{N}$ denotes the amount of localized charge  in the reference situation.

In this section, we consider noise in the dopant layer to be equivalent to a random change in the gate voltage $V_G$ with an amplitude $\delta V$.    If the charging energy
$(K_L/2)(e^*  + N_L  + \nu_{in} \phi - \gamma  V_G )^2$ for quasiparticles in the interior of the interferometer cell is close to a degeneracy point, this voltage fluctuation is large enough to change $N_L$ by one. It is possible that voltage fluctuations in the donor layer have a lever arm different from that of an external gate, but we will assume that the difference in lever arm is already included${}^\dagger$ in the value $\delta V$. In the range of gate voltages $V_G$ where a fluctation $\delta V$ leads to a change in $N_L$, the resulting phase jump in  is given by
%
\begin{equation}
\delta \theta = \theta_a  - 2 \pi {(e^*)^2 \over \Delta \nu} {K_{IL} \over K_I} + 2 \pi e^*\left(\beta + {K_{IL} \over K_I}
{\gamma \over \Delta \nu} \right) \delta V \ \ .
\label{phasejump.eq}
\end{equation}
%
The first contribution on the right hand side is the anyonic statistical phase, the second contribution describes the change in interferometer area due to the Coulomb coupling with the quasiparticle charge (we will denote it by $\delta \theta_C$ in the following), and the third term describes the change in
interferometer area as a direct response to the fluctuating voltage.

It is possible to give an upper limit for the magnitude of the Coulomb contribution to  phase jumps  if the
interferometer is in the Aharonov-Bohm regime for integer quantum Hall states as in \cite{Willett,Kang}. Then,  the phase jump due to Coulomb  coupling in the integer regime has to be smaller than $1/2$, hence we know that $K_{IL}/K_L < 1/2$. For the 7/3-state, we have $e^*=1/3$, $\Delta \nu = 1/3$, and hence $-\pi/3 < \delta \theta_C < 0$. This implies that $|\delta \theta_C|$ is at most half as big as the anyonic $\theta_a = 2 \pi/3$.
For the 5/2-state and interference of charge $e^*=1/4$ quasiparticles above an underlying $7/3$-state (implying $\Delta \nu = 1/6$), we find
$-3 \pi/8 < \delta \theta_C < 0$, such that $|\delta \theta_C|$  can be quite large as compared to the anyonic $\theta_a = \pi/4$.

An estimate for the magnitude of $\delta V $ can be obtained from the width of the interval of gate voltages $[V_{G,{\rm min}}, V_{G,{\rm max}}]$ over which phase jumps can be observed, as $\delta V = V_{G,{\rm max}} - V_{G,{\rm min}}$ in the low temperature limit. Under the assumption that the interior of the interferometer is close to a quantized Hall state and that quasiparticles are dilute, one has $\gamma \ll \beta$ and can neglect the term proportional to $\gamma$ in Eq.~(\ref{phasejump.eq}). As $\beta$ is known from the evolution of the interference phase in regions without telegraph noise, the relative strength $K_{I,L}/K_I$ of Coulomb coupling
between interior and edge
can be determined  from the jump $\delta \theta_{\rm C,  integer}$  of the interference phase in the integer regime as
$K_{IL}/K_I = - (\delta \theta_{\rm C, integer}/2 \pi) + \beta \delta V$. Then, under the assumption that
$\gamma \ll \beta$, both the Coulomb and direct contribution to the phase jumps can be predicted from Eq.~(\ref{phasejump.eq}) for fractional quantum Hall states.

\vspace*{5pt}

\noindent\begin{minipage}{\textwidth} \small ${}^\dagger$ Note that in principle, the lever arm for fluctuations in the donor layer can deviate from the lever arm for an external gate by different amounts depending on whether one considers coupling to the interferometer area or coupling to the number of localized quasiparticles in the interior of the interferometer. If this is the case, we assume that the
the lever arm for coupling to quasiparticles in the interior is included in the definition of $\delta V$, and that a parameter  $\tilde{\beta}$ is needed to describe the influence on the area of the interferometer..
\end{minipage}

\newpage

\subsection*{Supplemental Material B:  Coupled Interferometer/Dopants Model}
\label{sec:charging}

In order  to discuss  corrections to the anyonic phase  due to the Coulomb interaction inside the interferometer, we analyze a classical charging energy.  The general scheme for calculation follows the discussion of Refs. \onlinecite{RosenowHalperin,Nederetal}. We define a reference situation for the interferometer, in which the area enclosed by the interfering edge states is $A_0$. A deviation $\delta A$ from this area gives rise to an additional charge (measured in units of the electron charge)
%
\begin{equation}
\delta n_I  = \Delta \nu  {B_0 \delta A  \over \Phi_0}
\end{equation}
%
on the edge. Here, $B_0$ denotes the magnetic field in the reference situation, and $\Delta \nu = \nu_{\rm in} - \nu_{\rm out}$ is the difference of filling fractions $\nu_{\rm in}$ of the  quantized Hall state inside the interferometer and the
filling fraction $\nu_{\rm out}$ of the underlying quantized Hall state outside the interferometer. The  interference phase in the presence of  this charge  on the interfering edge is given by
%
\begin{equation}
\theta   =   2 \pi e^*(\phi  + \beta V_G) \  +\  N_L  \theta_a \ + \ 2 \pi  {e^* \over \Delta \nu}  \delta n_I  \ .
\label{phase.eq}
\end{equation}
%
Here, $\phi = (B - B_0) A_0 / \Phi_0$ denotes the change of flux through the reference area in units of the flux quantum, and $N_L$ denotes the number of localized quasiparticles. In order to determine the influence of charging effects on the interference phase, we parameterize the charging energy of the  combined system of interferometer and a fluctuating two-level system as
%
\begin{eqnarray}
E & = & {K_I \over 2} (\delta n_I)^2 \ + \  {K_L \over 2} (\delta n_L)^2\  +\  {K_D\over 2} (\delta S)^2
\label{energy.eq}\\
& & + K_{I,L} \delta n_I \delta n_L  \ + \ K_{D,I} \delta S \delta n_I \ + \ K_{D,L} \delta S \delta n_L \nonumber
\end{eqnarray}
%
Here, we assume that the fluctuating two-level system in the donor layer can take the values $S = \pm 1$, and
that $\delta S = S - S_0$ where $S_0$ is a charge offset. The variable $\delta n_L$ is defined as
%
\begin{equation}
\delta n_L = e^* N_L + \nu_{\rm in} \phi - \gamma V_G \ \ .
\end{equation}
%
In this definition, both $N_L$ and $V_G$ are measured with respect to the reference situation, and
$\gamma = {\partial \overline{N} \over \partial V_G}$. Here, $\overline{N}$ denotes the amount of localized charge  in the reference situation.

As $\delta n_I$ is a continuous variable, it assumes the value which minimizes the total charging energy, hence
%
\begin{equation}
\delta n_I \ = \ - {1 \over K_I} ( K_{D,I} \delta S \ + \ K_{I,L} \delta n_L)  \ \ .
\label{deltanI.eq}
\end{equation}
%
Inserting this  expression   in the effective charging energy Eq.~(\ref{energy.eq}), we find
%
\begin{equation}
E \ = \ {1 \over 2} \delta n_L^2 \left(K_L - {K_{I,L}^2 \over K_I} \right) \ + \ {1 \over 2} (\delta S )^2
\left( K_D - {K_{D,I}^2 \over K_I} \right) \ + \ \delta n_L \delta S \left( K_{D,L} -
{K_{D,I} K_{I,L} \over K_I} \right) \ \ .
\end{equation}
%
We now consider a scenario where the fluctuating two level system in the donor layer changes from
$S= +1$ to $S = -1$, and at the same time  the excess charge in the interior of the interferometer changes from $\delta n_{L,0}$ to $\delta n_{L,0} + e^*$.
 Then, the difference in $\delta n_I$ is given by
%
\begin{equation}
\delta n_I(S=+1) - \delta n_I(S=-1) \ = \ {1 \over K_I} \left( 2 K_{D,I} - e^* K_{I,L} \right) \ \ ,
\end{equation}
%
%
For the same process, the change in energy is given by
%
\begin{equation}
\label{eq:changeinenergy}
\Delta E \ = \ \left\{  \delta n_{L,0} - e^*/2\right\} \left[ \left( e^* K_L - 2 K_{D,L}\right)   - {K_{I,L} \over K_I} \left( e^* K_{I,L} - 2 K_{D,I}
 \right) \right]
\end{equation}
%
The first two  terms in the square bracket describe the energy cost for moving a charge from the dopant layer to the interior of the interferometer, whereas the second two terms describe the difference in Coulomb coupling strength between  dopant layer and interfering  edge on the one hand and  between the interior of the interferometer and the interfering edge on the other hand. We note that for thermally activated charge fluctuations in the dopant layer, $\Delta E$ should  not be larger than $k_B T$. As typical charging energies for micron sized
interferometers are on the scale of Kelvins, and experiments are performed on the scale of 10 mK,
this implies that  fluctuations occur only if either i) the prefactor is small or ii) the term in the square bracket  is much smaller than the natural energy scale. As the size of the prefactor can be varied  by varying the magnetic field, possibility i) can be excluded experimentally.   We thus consider the possibility ii).  If both terms in round brackets are  small individually, then we can immediately conclude that $\delta n_I$ and hence $\delta \theta_C$ are small. We now argue that this should indeed be the case: under the assumption that the dopant layer and the interior of the interferometer are spatially close to each other, we expect the coupling strengths of the dopants to either the edge or to the localized states will be similar to the coupling strength of the localized states to the edge or the localized states.  More specifically we expect that the ratio of the $K_{D,I}$ to $K_{I,L}$ should be similar to the ratio of $K_{D,L}$ to $K_L$.   Neglecting for the moment that the nominal charge fluctuation in the dopant layer is 2 whereas the charge fluctuation of the localized states is $e^*$, the equality of the two ratios implies that the difference between the two round brackets can only be small when the ratio $K_{I,L}^2/(K_I K_L) \approx 1$ such that the interferometer is in the extreme CD limit, where there is almost a degeneracy in electrostatic energy for moving  localized charges to the edge.
Vice versa, if the interferometer is not in the extreme CD limit, then it is unlikely that the square bracket is small due to a cancellation between the round brackets, and instead the two terms in round brackets have to be small individually.    We thus expect that in order to have $\Delta E \approx k_B T$ (such that fluctuations occur) we should have $(2 K_{D,I} - e^* K_{I,L})/K_I \ll1$ small such that $\delta n_I \ll 1$.  Hence we find
 that the Coulomb correction to the anyonic phase
%
\begin{equation}
\delta \theta_C \ = \ 2 \pi {e^* \over \Delta \nu} {1 \over K_I} (2 K_{D,I} - e^* K_{I,L}) \ \ .
\end{equation}
%
is small as well. We would like to mention that doping layers in heavily doped samples are believed to have a glassy dynamics, such that the occurrence of self-organized low energy charge fluctuations with small $\delta \theta_C$ can be expected. In addition, there is no reason that the charge fluctuations in the dopant layer must have   one electron charge exactly. Due to screening by other charges in the dopant layer arbitrary fractions of the electron charge can be realized, and we expect that for low energy fluctuations the effective fluctuating charge should be close to the charge $e^*$ of localized charges.


\begin{references}
\bibitem{commentonising} Strictly speaking the braiding statistics of quasiparticles the 5/2 state are expected to be Ising up to an
    abelian phase.
\bibitem{endnote1} The sign of of the phase slip may be positive or negative depending on whether quasiparticles or holes are added to the dot as the gate voltage is swept.  This sign may not be directly related to the sign of the gate voltage depending on details such as whether the gate is an area gate or a density gate.
\bibitem{endnote2} For a set of excitation energies $E_n$ the visibility of interference scales as $\prod_n \tanh(E_n/2 T)$.  If the energy levels are equally spaced, the visibility (in the case of even number of quasiparticles in the cavity) is reduced to below half at $\Delta E=T$, is 1/10 at twice this temperature, and drops drastically at higher temperatures.
\end{references}
\end{document}